\documentclass[prl,twocolumn,amsmath,showpacs,amssymb,floatfix]{revtex4}
\def\figwidth{0.9\linewidth}
\usepackage[final]{graphicx}
\def\A{{\text{A}}}
\def\B{{\text{B}}}
\def\E{{\text{E}}}
\def\etal{\textit{et~al.}}

\begin{document}
\title{Mixing effects for the structural relaxation
       in binary hard-sphere liquids}
\date{\today}
\def\tumphy{%
  \altaffiliation[On leave from: ]{Physik-Department, %
  Technische Universit\"at M\"unchen, 85747 Garching, Germany}}
\author{G.~Foffi}
\author{W.~G\"otze}\tumphy
\author{F.~Sciortino}
\author{P.~Tartaglia}
\author{Th.~Voigtmann}\tumphy
\affiliation{Dipartimento di Fisica and INFM Center for Statistical Mechanics
             and Complexity, Universit\`a di Roma ``La Sapienza'', Piazzale
             Aldo Moro 2, 00185 Roma, Italy}

\begin{abstract}
We report extensive molecular-dynamics simulation results for binary mixtures
of hard spheres for different size disparities and different mixing percentages,
for packing fractions up to $0.605$ and over a characteristic time interval
spanning up to five orders in magnitude. We explore the changes in the
evolution of glassy dynamics due to mixing and discover two
opposite scenarios: for large size disparity, increasing the mixing
percentage of small particles leads to a speed-up of long-time
dynamics, while small disparity leads to a slowing down. These results agree with predictions based on the mode coupling theory for ideal glass transitions.
\end{abstract}
\pacs{64.70.Pf, 82.70.Dd}

\maketitle

This paper deals with structural relaxation in liquids, i.e.\ the dynamical
phenomena which are precursors of the glass transition. These processes are
characterized by rather complex dependencies of correlation functions on time and parameters like the density and the temperature. The introduction of
several new experimental techniques and great progress in molecular dynamics
simulation studies has unfolded a wealth of facts about the evolution of
structural relaxation upon cooling or compressing glass-forming
liquids \cite{crete,pisa}.

To establish an understanding of structural relaxation and the glass
transition, one should focus on systems as simple as possible. Atomic
one-component systems with conventional interaction potentials cannot be used
since they crystallize before structural relaxation is fully developed.
Beginning with the work of Hansen and collaborators \cite{Roux1989},
binary mixtures of simple particles were used for molecular dynamics
studies of the glass transition. A binary Lennard-Jones system has been used
extensively in recent years to analyze structural relaxation \cite{Kob2003prex}. 

In previous work, mixing was introduced merely as a means of suppressing
crystallization \cite{Henderson1996}. In the following we analyze systematically
the effect of mixing on structural relaxation to identify the influence of
composition changes and variation of the particle size disparity on the glassy
dynamics. We focus on binary hard-sphere mixtures (HSM), i.e.\ particles
interacting via an hard-core potential, which we study via extensive molecular
dynamics simulations. 

A specific motivation for our studies comes from the light-scattering data by
Williams and van~Megen for a HSM of colloidal particles \cite{Williams2001b}.
The system was prepared to approximate a binary HSM for the ratio
$\delta=0.60$ of the particle diameters.
Three mixing effects have been reported.
If the percentage of the smaller particles increases from $10\%$ to $20\%$
of the relative packing fraction, then (i) the time scale for the final decay
of the density correlators decreases; (ii) the plateau value for
intermediate times increases; (iii) the initial part of the structural
relaxation slows down. The first effects means that mixing has stabilized
the liquid as if the smaller particles provide some lubrification. This
effect has some analog in the plasticization observed in dense polymeric
liquids due to mixing with polymers of shorter lengths. However, the effects
(ii) and (iii) indicate a stiffening of the dynamics upon mixing. These
effects have never been reported for conventional systems, and one might wonder
whether these are structural relaxation phenomena rather than colloid-specific
features caused by, e.g., hydrodynamic interactions or polydispersity. Since
our simulation studies are done implementing a Newtonian microscopic dynamics,
this question will be answered by our results.

Another motivation of our work is provided by recent predictions for binary
HSM based on mode-coupling theory (MCT) \cite{Goetze1991b} calculations. This
theory, which allows first principle evaluations of density-fluctuation
correlators within a regime where structural relaxation dominates the dynamics,
explains the light-scattering data for slightly polydisperse hard-sphere
colloidal suspensions \cite{Megen1995}. Extending MCT to binary HSM, the three
mentioned mixing effects have been identified as structural relaxation
properties \cite{Goetze2003}. The light scattering data for the $\delta=0.60$
mixture could be described quantitatively to a certain extent
\cite{Voigtmann2003pre}. However, surprisingly, the theory predicts two
different scenarios. The speed-up of the dynamics reported in
Ref.~\cite{Williams2001b} was found only for sufficiently large size disparity,
say a size ratio $\delta\lesssim0.65$. For $\delta\gtrsim0.8$, the opposite
effect was predicted, i.e., mixing slows down the dynamics and the ideal-glass
critical packing fractions decreases. The data reported in this letter confirms
these astonishing predictions.

We perform standard constant-energy molecular dynamics \cite{Rapaport1997}
simulation for a binary mixture of 1237 hard-sphere particles. The two species
($\A$ and $\B$) have diameters $d_\A$ and $d_\B$ respectively, with
$d_\A\ge d_\B$. The masses of the two species are taken as equal, thus all
particles have the same thermal velocity, denoted as $v_{\text{th}}$. Units of
length and time are chosen such that $d_\A=1$ and $v_{\text{th}}=1$.
We use the size ratio $\delta=d_\B/d_\A$ as a control parameter specifying the
size disparity. To model a large size disparity, we choose $\delta=0.60$. 
To model systems of small size
disparity we study the value $\delta=0.83$. For each of the two $\delta$
values, we study two different values of the relative packing fraction of the
small species, $x=\varphi_\B/(\varphi_A+\varphi_B)$, where
$\varphi_\alpha=(\pi/6)\varrho_\alpha d_\alpha^3$. For the $\delta=0.60$
system, we have studied  $x=0.10$ and $x=0.20$, corresponding to a fraction of
$\B$ particles equal to  $34\%$ and $54\%$. For the $\delta=0.83$ system, we
have studied the cases $x=0.276$ and $x=0.37$, corresponding to having $40\%$
and $50\%$ of $\B$ particles. For each $\delta$ and $x$ values, we study
several values of the total packing fraction $\varphi=\varphi_\A+\varphi_\B$,
covering a region where dynamics slows down by 4 decades. During the long
simulation runs, we check for crystallization by monitoring the time
evolution of the pressure and by visual inspection of the configurations. We
also evaluate the structure factor to make sure that no crystalline peaks have
developed.

\begin{figure}[t]
\includegraphics[width=\figwidth]{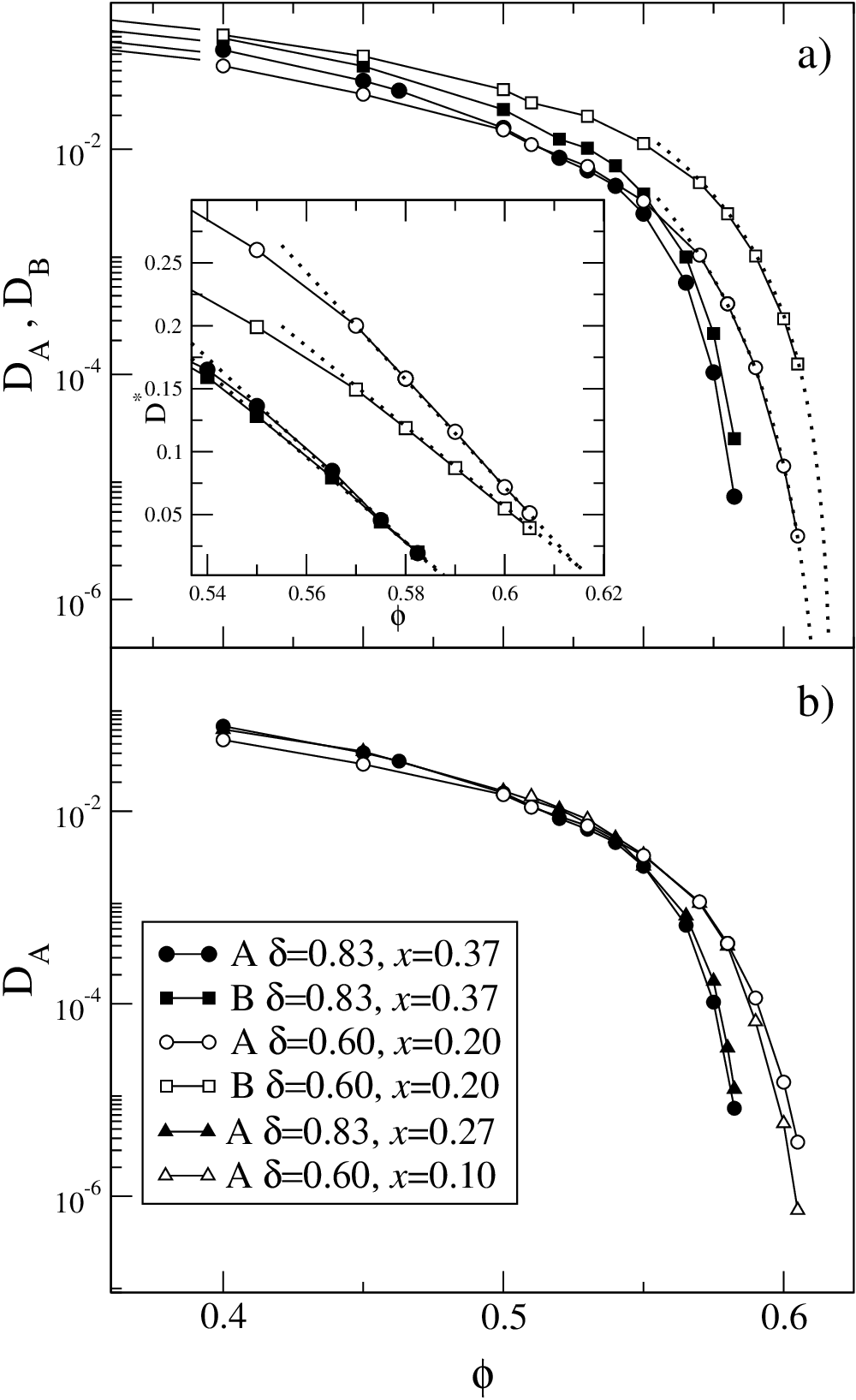}
\caption{\label{figdiff}
  Diffusivities $D_\alpha$, $\alpha=\A,\B$, for binary hard-sphere mixtures
  with size ratios $\delta=d_\B/d_\A=0.60$ (open symbols) and $\delta=0.83$
  (filled symbols); packing contributions of the smaller spheres
  $x=\varphi_\B/\varphi$ as in the legend. (a)
  Results for the two different $\delta$ with $\alpha=A$
  (circles) and $\alpha=B$ (squares).  The solid lines for $\varphi<0.39$ are the Enskog results for $D_\B(\delta=0.6)$, $D_\B(\delta=0.83)$, $D_\A(\delta=0.83)$,
and $D_\A(\delta=0.6)$, from top to bottom.
Dotted lines demonstrate power-law
  fits with exponents $\gamma_\alpha(\delta,x)$, see text for details.
  The inset shows $D_\alpha^*=D_\alpha^{1/\gamma_\alpha}$ to demonstrate the  extrapolation to zero diffusivity.
  (b) Variation of $D_\A$ upon changes of $x$ for both $\delta$ values;
  circles (triangles) refer to the higher (lower) $x$ value studied.
}
\end{figure}

The $\varphi$ dependence of the diffusion coefficient, evaluated from the long
time limit of the mean square displacement \cite{Boon1980}, is shown in
Fig.~\ref{figdiff}. Note that the variation of the diffusivities extends over
more than four decades. 

It is important to separate mixing effects already observed in normal liquid
states from mixing effects which are peculiar of the structural relaxation
dynamics. In hard-spheres, mixing effects for the normal-liquid dynamics are
qualitatively explained within Enskog's kinetic theory. The diffusion constant
$D$ of the hard-sphere system (HSS) is expressed in terms of Enskog's collision rate $\nu$,
$D^\E=3v_{\text{th}}^2/(2\nu)$. This rate modifies Boltzmann's collision rate
for dilute gases by the contact value $g_{\A\A}$ of the pair-distribution
function: $\nu=4\sqrt\pi v_{\text{th}}^2\varrho_\A g_{\A\A}d_\A^2$
\cite{Boon1980}. Enskog's formula is readily generalized to mixtures with
equal constituents' masses, giving
$D_\alpha^\E=3v_{\text{th}}^2/(2\nu_\alpha)$, with
$\nu_\alpha=4\sqrt{\pi}v_{\text{th}}^2\sum_\beta\varrho_\beta g_{\alpha\beta}
d_{\alpha\beta}^2$. Here $g_{\alpha\beta}$ denotes the pair-distribution
function of the colliding pair $\alpha$ and $\beta$ for the distance at
contact, $d_{\alpha\beta}=(d_\alpha+d_\beta)/2$.  
Results for $D_\alpha^\E$  are included in Fig.~\ref{figdiff} as lines for
$\varphi<0.39$, using values from Percus-Yevick theory for $g_{\alpha\beta}$.
The relevant normal-liquid mixing effects for the HSM, as given by the Enskog
theory, can be summarized in the following: (i) $D_\A$ decreases and $D_\B$
increases upon decreasing $\delta$; (ii) both $D_\A$ and $D_\B$ decrease upon
increasing $x$; (iii) all $D_\A$ values are smaller than all the $D_\B$ values.
A detailed discussion of normal mixing effects for an approximation
to HSM has been published in Ref.~\cite{Schaink1992}.

\begin{figure}
\includegraphics[width=\figwidth]{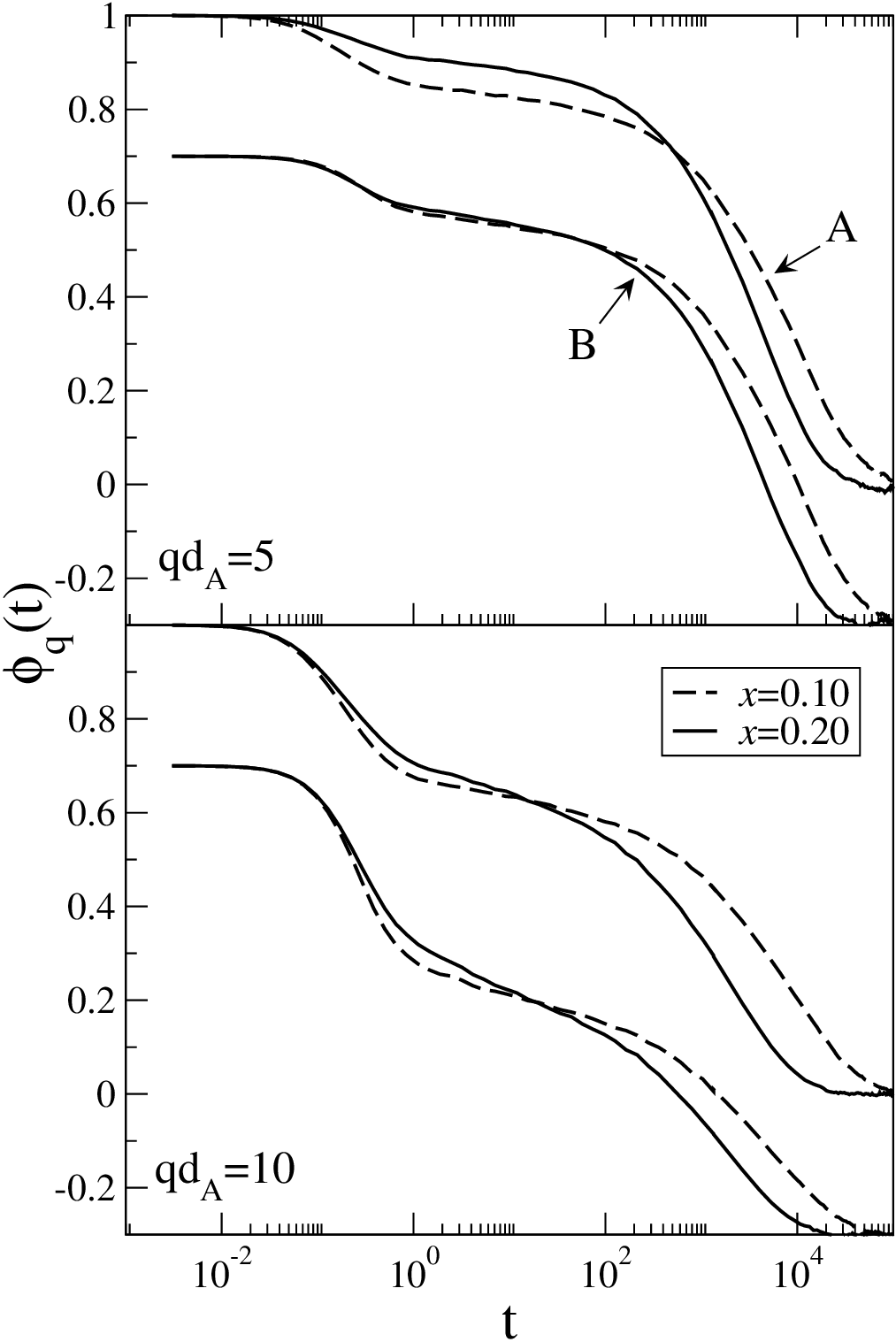}
\caption{\label{figphi1}
  Normalized density-correlation functions for binary hard-sphere mixtures
  at a total packing fraction $\varphi=0.600$ for particles with diameter
  ratio $\delta=0.60$. The full and dashed lines present mixtures with relative
  packing contributions of the smaller particles $\B$ of $x=0.20$ and $x=0.10$,
  respectively. Upper and lower panel exhibit the results for wave vectors
  $qd_A=5$ and $qd_A=10$, respectively. The results for the particles
  $\B$ have been shifted down by $0.3$ relative to the ones for the
  particles $\A$.
}
\end{figure}

In agreement with the pioneer work of Alder \etal\ for the HSS \cite{Alder1970b}, the data
for $D_\alpha$ are described correctly on a $30\%$ accuracy level by the Enskog
theory for $\varphi<0.40$. Upon increasing $\varphi$ beyond $0.40$, the cage
effect starts dominating and leads to a suppression of $D$ and $D_\alpha$ below
the corresponding Enskog values. Still we find that up to the large value
$\varphi=0.48$ there are only normal-liquid mixing effects since the
$\log D_\alpha$-versus-$\varphi$ curves do not exhibit crossings.

Increasing $\varphi$ further, the curves in Fig.~\ref{figdiff} get a stronger
bending, i.e., the cage-effect enhances. If $\varphi$ reaches $0.56$, the
diffusivities are about one order of magnitude smaller than in a conventional
liquid. Within the interval $0.48<\varphi<0.56$, there occur crossings of the
curves, i.e. the mixing properties change qualitatively. The crossing observed
in Fig.~\ref{figdiff} demonstrates a first remarkable feature of structural
relaxation, viz., differently from Enskog predictions, diffusion of $\B$
particles for $\delta=0.83$ becomes smaller than $D_\A$ for the
$\delta=0.60$ case. For $\varphi>0.56$, a new pattern evolves, showing the
mixing effects for the fully developed structural relaxation. To support the
association of the slowing down of the dynamics to the approach towards a
glass transition, the diffusivities have been fitted by a power-law function,
$D_\alpha=\Gamma_\alpha(\delta,x)
[\varphi_\alpha^c(\delta,x)-\varphi]^{\gamma_\alpha(\delta,x)}$.
As shown in Fig.~\ref{figdiff}, the power-law fit can account well for the
diffusivity changes over more than two orders of magnitude. For the four
mixtures studied, we find that the critical packing fraction of species
$\A$ deviates from that of species $\B$ by not more than $0.0008$, so that
we can confidently conclude, in agreement with MCT predictions, that each
mixture exhibits a single critical packing fraction $\varphi^c(\delta,x)$.
In the asymptotic limit $\varphi\to\varphi^c(\delta,x)$, the MCT exponent
$\gamma$ should be the same for $D_\A$ and $D_\B$. This is not the case for our
fit results, as also found in previous studies
\cite{Kob1994}, possibly because of preasymptotic correction effects.

\begin{figure}[t]
\includegraphics[width=\figwidth]{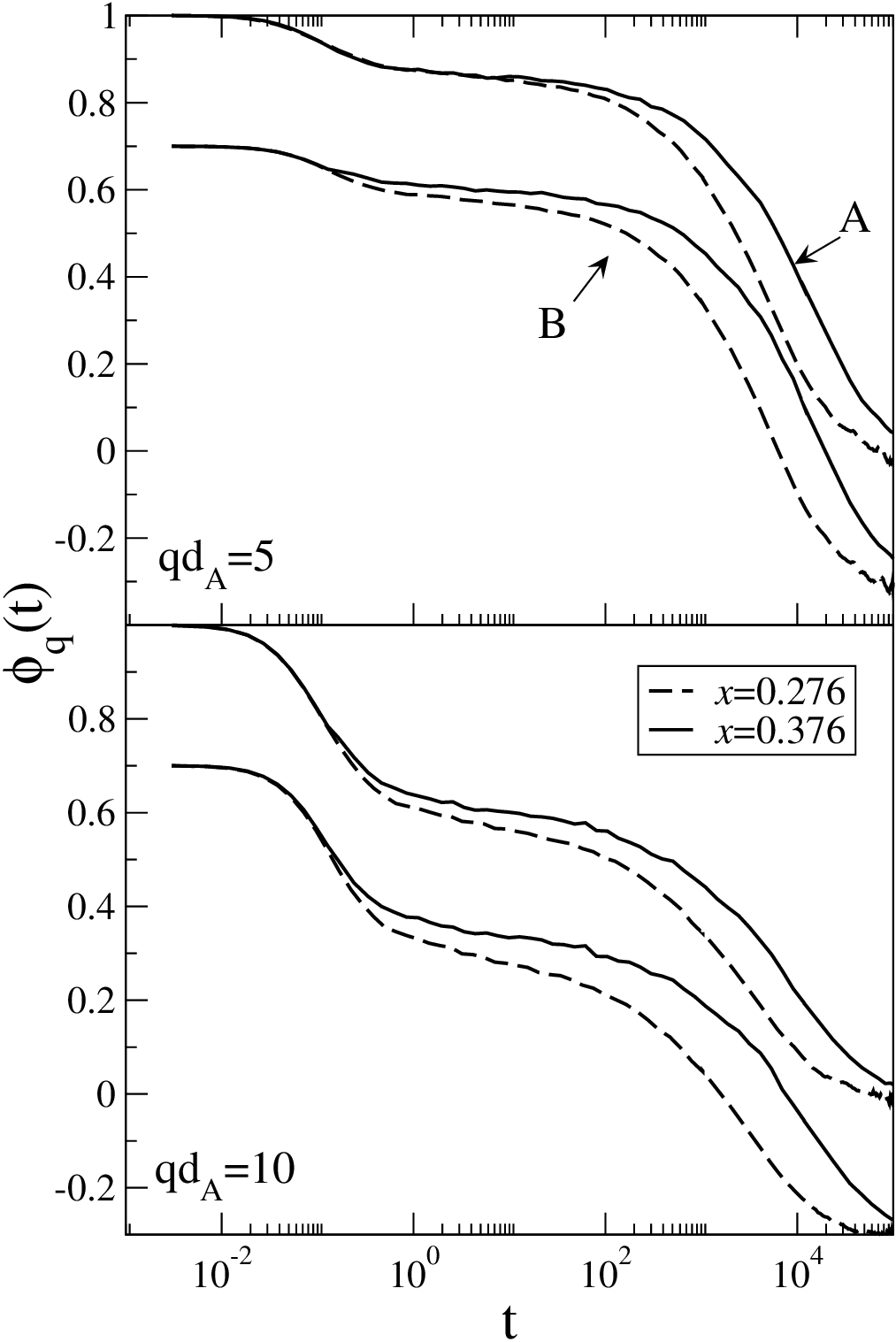}
\caption{\label{figphi2}
  As in Fig.~\protect\ref{figphi1}, but for a packing fraction
  $\varphi=0.582$, size ratio $\delta=0.83$, and relative packing contributions
  of small particles $x=0.37$ (full lines) and $x=0.276$ (dashed lines).
}
\end{figure}

The presence of crossings in the diffusivity curves on entering the region
where structural relaxation becomes relevant clearly shows that dynamics for
$\delta=0.83$ is significantly slower than the one for $\delta=0.60$.
Moreover---differently from the regime of normal-liquid dynamics where the
diffusivities $D_\alpha(\delta,x)$ exhibit the same trend with changes of $x$
for $\delta=0.83$ and $\delta=0.60$---there is a qualitatively different
$x$-dependence of the long-time dynamics for small and large size disparity,
as recently predicted by MCT \cite{Goetze2003}. As shown in
Fig.~\ref{figdiff}b, while at $\delta=0.83$ dynamics becomes slower on
increasing $x$, the opposite behavior is observed at $\delta=0.60$.
This second remarkable feature---the plasticization phenomenon alluded to
before and exhibited by the $\delta=0.60$ simulation data---is clearly 
shown in Fig.\ref{figdiff}b. Consistently with these findings, at
$\delta=0.60$, the critical packing fraction increases on going from $x=0.10$
($\varphi^c(x=0.10)=0.6139\pm0.0004$) to $x=0.20$
($\varphi^c(x=0.20)=0.6169\pm0.0004$). The critical packing fractions for the
system with $\delta=0.83$ show the opposite trend: they decrease on increasing
$x$ from $x=0.276$ ($\varphi^c(x=0.276)=0.5881\pm0.0004$) to $x=0.37$
($\varphi^c(x=0.37)=0.5877\pm0.0004$). This third remarkable finding means that
mixing stabilizes the glass for the system with small size disparity and
destabilizes it for large size disparities. Different from the plasticization
effect discovered by Williams and van~Megen for $\delta=0.60$, for
$\delta=0.83$ an increase of the concentration of the smaller minority
particles leads to a slowing down of the long-time density-fluctuation
dynamics.
The different scenarios for the long-time relaxation scales are demonstrated also in Figures \ref{figphi1} and \ref{figphi2}. They show the $x$ dependence of the
density auto-correlation functions $\phi_\alpha(q,t)$ for $\alpha=\A$
and $\B$ at two different wave vectors, below ($qd_\A=5$) and above ($qd_\A=10$)
the first peak of the structure factor $(qd_\A=7$), respectively. 

Within MCT, the ideal glass states are characterized by arrest of the density
fluctuations. Within the liquid state, for $\varphi$ below but close to
$\varphi^c$, the $\phi_\alpha(q,t)$-versus-$\log t$ curves exhibit a plateau at
$f_\alpha^c(q)$, the so-called critical Debye-Waller factor. The curves are the
flatter the smaller $\varphi^c-\varphi$, and the length of the plateau
increases. This plateau, an outstanding feature of structural relaxation,
is clearly exhibited in Figs.~\ref{figphi1} and \ref{figphi2}. The whole
structural-relaxation process consists of two steps. The first step deals with
the relaxation towards the plateau $f_\alpha^c(q)$. The second step is the
long-time process dealing with the decay of $\phi_\alpha(q,t)$ from the
plateau to zero. The three mixing features discussed above concerned the time
scale of the second relaxation step. 

The results for $\phi_\A(q,t)$ shown in Figs.~\ref{figphi1} and \ref{figphi2}
for $qd_\A=5$ exhibit a fourth remarkable mixing effect: upon increasing $x$,
the plateaus of the correlators increase. Accompanied with this is a flattening
of the $\phi_\A(q,t)$-versus-$\log t$ curve in the plateau region. An
indication of the same effect is also shown by the $\phi_\B(q,t)$. These
findings are in accord with the two mixing anomalies discovered in
the experimental study of the colloidal $\delta=0.60$ system
\cite{Williams2001b}. Our results show that they are
structural-relaxation phenomena that occur also in systems with Newtonian
dynamics, and that they occur for small as well as for large size disparity
of the particles. We also note that on increasing $x$ at $\delta=0.6$,
the increase of the height of the plateau value, associated with the speed-up
of the dynamics, forces the relaxation curves for the same $\varphi$ but for
different $x$ to cross at the beginning of the second relaxation step.
For $\delta=0.83$, an increase of $x$ at fixed $\varphi$ generates a slower
dynamics and hence there is no crossing of the curves in the time interval for
the second relaxation step.

To conclude, we have studied structural relaxation of four binary hard-sphere
mixtures via molecular dynamics simulations over dynamic ranges extending over
about five orders of magnitude. Surprising mixing effects for the slow
dynamics have been identified. In particular, we have shown that increasing the
mixing percentage of the smaller minority particles can lead to a speeding up
as well as to a slowing down of the long-time decay processes, depending on
whether the size disparity is large or small, respectively. There is also an
increase of the plateau of the density-autocorrelation functions for small and
intermediate wave vectors reflecting a stiffening of the nearly arrested glass
structure. These findings, which pose a challenge to theories of the glass
transition, show in particular, that the description of a glass-forming mixture
by an effective one-component liquid cannot be possible for all properties of
interest. The reported results also confirm the conclusions arrived
at in a light-scattering study of a quasi-binary colloidal
suspension \cite{Williams2001b}, and predictions within
mode-coupling theory \cite{Goetze2003}. Finally, our work suggests that mode-coupling theory
can contribute to an understanding of qualitative trends for the
microscopic details in the dynamics of glass-forming liquids.

\begin{acknowledgments}
W.G.\ and Th.V.\ thank their colleagues from the University of Rome for
their kind hospitality during the time this work was performed. Our
collaboration was supported by DYGLAGEMEM and the Deutsche Forschungsgemeinschaft through DFG grant Go 154/12-1.
G.F., F.S.\ and P.T.\ acknowledge support from MIUR Prin and Firb and
INFM Pra-Genfdt. We thank S.~Buldyrev for providing us the MD code
for HSM.
\end{acknowledgments}

\bibliography{mct,add}

\begin{thebibliography}{17}
\expandafter\ifx\csname natexlab\endcsname\relax\def\natexlab#1{#1}\fi
\expandafter\ifx\csname bibnamefont\endcsname\relax
  \def\bibnamefont#1{#1}\fi
\expandafter\ifx\csname bibfnamefont\endcsname\relax
  \def\bibfnamefont#1{#1}\fi
\expandafter\ifx\csname citenamefont\endcsname\relax
  \def\citenamefont#1{#1}\fi
\expandafter\ifx\csname url\endcsname\relax
  \def\url#1{\texttt{#1}}\fi
\expandafter\ifx\csname urlprefix\endcsname\relax\def\urlprefix{URL }\fi
\providecommand{\bibinfo}[2]{#2}
\providecommand{\eprint}[2][]{\url{#2}}

\bibitem[{cre(2002)}]{crete}
\emph{\bibinfo{title}{Proceedings of the 4th International Discussion Meeting
  on Slow Relaxations in Complex Systems}}, {J}.~Non-Cryst.~Solids,
  \textbf{307--310} (\bibinfo{year}{2002}).

\bibitem[{pis(2003)}]{pisa}
\emph{\bibinfo{title}{Proceedings of the 3rd Workshop on Non-Equilibrium
  Phenomena in Supercooled Fluids, Glasses, and Amorphous Materials}},
  {J}.~Phys.: Condens.~Matter, \textbf{15}(11) (\bibinfo{year}{2003}).

\bibitem[{\citenamefont{Roux et~al.}(1989)\citenamefont{Roux, Barrat, and
  Hansen}}]{Roux1989}
\bibinfo{author}{\bibfnamefont{J.~N.} \bibnamefont{Roux}},
  \bibinfo{author}{\bibfnamefont{J.-L.} \bibnamefont{Barrat}},
  \bibnamefont{and} \bibinfo{author}{\bibfnamefont{J.-P.}
  \bibnamefont{Hansen}}, \bibinfo{journal}{J.~Phys.: Condens.~Matter}
  \textbf{\bibinfo{volume}{1}}, \bibinfo{pages}{7171} (\bibinfo{year}{1989}).

\bibitem[{\citenamefont{Kob}(2003)}]{Kob2003prex}
\bibinfo{author}{\bibfnamefont{W.}~\bibnamefont{Kob}} (\bibinfo{year}{2003}),
  \bibinfo{note}{cond-mat/0212344, \textit{Les Houches Summer School of
  Theoretical Physics, Session LXXVII}, in print}.

\bibitem[{\citenamefont{Henderson et~al.}(1996)\citenamefont{Henderson,
  Mortensen, Underwood, and van Megen}}]{Henderson1996}
\bibinfo{author}{\bibfnamefont{S.~I.} \bibnamefont{Henderson}},
  \bibinfo{author}{\bibfnamefont{T.~C.} \bibnamefont{Mortensen}},
  \bibinfo{author}{\bibfnamefont{G.~M.} \bibnamefont{Underwood}},
  \bibnamefont{and} \bibinfo{author}{\bibfnamefont{W.}~\bibnamefont{van
  Megen}}, \bibinfo{journal}{Physica A} \textbf{\bibinfo{volume}{233}},
  \bibinfo{pages}{102} (\bibinfo{year}{1996}).

\bibitem[{\citenamefont{Williams and van Megen}(2001)}]{Williams2001b}
\bibinfo{author}{\bibfnamefont{S.~R.} \bibnamefont{Williams}} \bibnamefont{and}
  \bibinfo{author}{\bibfnamefont{W.}~\bibnamefont{van Megen}},
  \bibinfo{journal}{Phys.~Rev.~E} \textbf{\bibinfo{volume}{64}},
  \bibinfo{pages}{041502} (\bibinfo{year}{2001}).

\bibitem[{\citenamefont{G\"otze}(1991)}]{Goetze1991b}
\bibinfo{author}{\bibfnamefont{W.}~\bibnamefont{G\"otze}}, in
  \emph{\bibinfo{booktitle}{Liquids, Freezing and Glass Transition}}, edited by
  \bibinfo{editor}{\bibfnamefont{J.~P.} \bibnamefont{Hansen}},
  \bibinfo{editor}{\bibfnamefont{D.}~\bibnamefont{Levesque}}, \bibnamefont{and}
  \bibinfo{editor}{\bibfnamefont{J.}~\bibnamefont{Zinn-Justin}}
  (\bibinfo{publisher}{North Holland}, \bibinfo{address}{Amsterdam},
  \bibinfo{year}{1991}), Les Houches Summer Schools of Theoretical Physics, pp.
  \bibinfo{pages}{287--503}.

\bibitem[{\citenamefont{van Megen}(1995)}]{Megen1995}
\bibinfo{author}{\bibfnamefont{W.}~\bibnamefont{van Megen}},
  \bibinfo{journal}{Transp.~Theory~Stat.~Phys.} \textbf{\bibinfo{volume}{24}},
  \bibinfo{pages}{1017} (\bibinfo{year}{1995}).

\bibitem[{\citenamefont{G\"otze and $\mbox{Th}$. Voigtmann}(2003)}]{Goetze2003}
\bibinfo{author}{\bibfnamefont{W.}~\bibnamefont{G\"otze}} \bibnamefont{and}
  \bibinfo{author}{\bibnamefont{$\mbox{Th}$. Voigtmann}},
  \bibinfo{journal}{Phys.~Rev.~E} \textbf{\bibinfo{volume}{67}},
  \bibinfo{pages}{021502} (\bibinfo{year}{2003}).

\bibitem[{\citenamefont{$\mbox{Th}$. Voigtmann}(2003)}]{Voigtmann2003pre}
\bibinfo{author}{\bibnamefont{$\mbox{Th}$. Voigtmann}} (\bibinfo{year}{2003}),
  \bibinfo{note}{to be published}.

\bibitem[{\citenamefont{Rapaport}(1997)}]{Rapaport1997}
\bibinfo{author}{\bibfnamefont{D.~C.} \bibnamefont{Rapaport}},
  \emph{\bibinfo{title}{The Art of Molecular Dynamics Simulation}}
  (\bibinfo{publisher}{Cambridge University Press},
  \bibinfo{address}{Cambridge, UK}, \bibinfo{year}{1997}).

\bibitem[{\citenamefont{Boon and Yip}(1980)}]{Boon1980}
\bibinfo{author}{\bibfnamefont{J.-P.} \bibnamefont{Boon}} \bibnamefont{and}
  \bibinfo{author}{\bibfnamefont{S.}~\bibnamefont{Yip}},
  \emph{\bibinfo{title}{Molecular Hydrodynamics}}
  (\bibinfo{publisher}{McGraw-Hill}, \bibinfo{address}{New York},
  \bibinfo{year}{1980}).

\bibitem[{\citenamefont{Schaink and Hoheisel}(1992)}]{Schaink1992}
\bibinfo{author}{\bibfnamefont{H.~M.} \bibnamefont{Schaink}} \bibnamefont{and}
  \bibinfo{author}{\bibfnamefont{C.}~\bibnamefont{Hoheisel}},
  \bibinfo{journal}{Phys.~Rev.~A} \textbf{\bibinfo{volume}{45}},
  \bibinfo{pages}{8559} (\bibinfo{year}{1992}).

\bibitem[{\citenamefont{Alder et~al.}(1970)\citenamefont{Alder, Gass, and
  Wainwright}}]{Alder1970b}
\bibinfo{author}{\bibfnamefont{B.~J.} \bibnamefont{Alder}},
  \bibinfo{author}{\bibfnamefont{D.~M.} \bibnamefont{Gass}}, \bibnamefont{and}
  \bibinfo{author}{\bibfnamefont{T.~E.} \bibnamefont{Wainwright}},
  \bibinfo{journal}{J.~Chem.~Phys.} \textbf{\bibinfo{volume}{53}},
  \bibinfo{pages}{3813} (\bibinfo{year}{1970}).

\bibitem[{\citenamefont{Kob and Andersen}(1994)}]{Kob1994}
\bibinfo{author}{\bibfnamefont{W.}~\bibnamefont{Kob}} \bibnamefont{and}
  \bibinfo{author}{\bibfnamefont{H.~C.} \bibnamefont{Andersen}},
  \bibinfo{journal}{Phys.~Rev.~Lett.} \textbf{\bibinfo{volume}{73}},
  \bibinfo{pages}{1376} (\bibinfo{year}{1994}).

\end{thebibliography}
\bibliographystyle{apsrev}

\end{document}